\documentstyle[eqsecnum,aps]{revtex}
\begin{document}
\draft
\title{Prolongation Approach to B\"{a}cklund Transformation of\\
 Zhiber-Mikhailov-Shabat Equation}
\author{Huan-xiong Yang }
\address{
China Center of Advanced Science and Technology
(World Lab.), P.O. Box 8730 Beijing 100080, China and\\
\footnote{mailing address}${}^{2}$ Department of Physics,
Hangzhou University, Hangzhou, 310028, P.R. China}
\author{You-Quan Li}
\address{Zhejiang Institute of Modern Physics, Zhejiang University,\\
Hangzhou,310027, P.R. China}
\date{Received ???}
\maketitle
\begin{abstract}
The prolongation structure of Zhiber-Mikhailov-Shabat (ZMS) equation
is studied by using Wahlquist-Estabrook's method. The Lax-pair for
ZMS equation and Riccati equations for pseudopotentials are formulated
respectively from linear and nonlinear realizations of the prolongation
structure. Based on nonlinear realization of the prolongation structure,
an auto-B$\ddot{a}$cklund transformation of ZMS equation is obtained.
\end{abstract}
\,\,\,\,
\pacs{PACS number: 11.30.Na; 02.20.Tw}


\section{Introduction}

  The off-conformally integrable models in two-dimensional space-time
have some common features. They have spectrum-dependent Lax pairs, an
infinite number of conserved currents and the underlying nonlinear
symmetries, and can be solved by means of inverse scattering method.
The $a^{(2)}_{2}$ Toda model, $i.\,e.$ Zhiber-Mikhailov-Shabat (ZMS)
model, is of such a fascinating class of integrable two-dimensional
field theories. It is the third and the last relativistic single scalar
Toda model (the others are Liouville and sine-Gordon models)\cite{Dodd,ZS},
and has significant applications in physical context. To our knowledge,
the equation of motion of ZMS model (ZMS equation) governs the propagation
of resonant ultra-short plane wave optical pulses in certain {\it degenerate}
media\cite{Lamb}. The integrability of ZMS model has been confirmed for a
long time. The soliton solutions of the ZMS equation by means of inverse
scattering method were given in\cite{Mikhailov}. The $S$-matrix
approach to the quantum version of the model was seriously investigated
by Izergin and Korepin\cite{Korepin} in terms of the quantum inverse
scattering method, and by Smirnov\cite{Smirnov} and Efthimiou\cite{Efthimiou}
in the framework of perturbative conformal field theory. Recently, we have
studied the infinitesimal dressing transformations and Lie-Poisson
structure hidden in ZMS model\cite{YCS}. Nevertheless, there would seem
not to have a good knowledge of the finite nonlinear symmetries (such as
the dressing group symmetry and B$\ddot{a}$cklund transformation) of the
model. In order to fill in the gap, we study the prolongation structure
of ZMS equation in the spirit of Wahlquist-Estabrook's(W-E's) prolongation
approach\cite{WE,Omote} in the present paper. Owing to the so-called
prolongation structure, we discover an auto-B$\ddot{a}$cklund
transformation of ZMS equation, which turns out to be a set of
Riccati-type differential equations of the so-called pseudopotentials.

\section{\bf Prolongation structure of ZMS equation}

W-E's prolongation structure is a very useful medium for searching
B$\ddot{a}$cklund transformation of nonlinear differential equations.
To obtain the prolongation structure of ZMS equation:
\begin{equation}
\label{e2.1}
\partial_{+}\partial_{-}\phi - 2 ( e^{\phi} - e^{-2\phi}) =0,
\end{equation}
we define the following 2-forms on a four-dimensional manifold with
coordinates associated with Eq.(\ref{e2.1}),
\begin{equation}
\label{e2.2}
\left\{ \begin{array}{l}
\alpha_{1}=d\phi \wedge dx^{+} - \pi_{\phi} dx^{-} \wedge dx^{+}, \\
\alpha_{2}=d\pi_{\phi} \wedge dx^{-}
+ 2 (e^{\phi} - e^{-2\phi}) dx^{-} \wedge dx^{+}.
\end{array}
\right.
\end{equation}
These 2-forms constitute a closed ideal and would become null on the
solution manifold $(\phi(x^{+},\,x^{-}),\,\pi_{\phi}(x^{+},\,x^{-}),\,
x^{+},\,x^{-})$. They are the Pfaff forms of ZMS equation. For the above
Pfaff forms we will assume that prolongation forms can be given by some
1-forms $\Omega^{a}~ (a=1,~2,~3,\cdots,N)$,
\begin{equation}
\label{e2.3}
\Omega^{a}=-dq^{a} + F^{a}(\phi, ~\pi_{\phi}, ~q)dx^{-}
+ G^{a}(\phi, ~\pi_{\phi}, ~q)dx^{+},
\end{equation}
where $N$ is an outstanding integer, $q^{a}$ are the so-called
pseudopotentials, and $F^{a}$ and $G^{a}$ are functionals of fields
$\pi_{\phi}$, ~$\phi$ and $q^{a}$.

The concept of pseudopotential plays a crucial role in the discussions of
B$\ddot{a}$cklund transformations and Lax pairs in W-E's prolongation
method. As a matter of fact, the expected B$\ddot{a}$cklund transformation
and the first-order differential equations satisfied by Lax pair of ZMS
Eq.(\ref{e2.1}) will be formulated as the differential equations for suitably
defined pseudopotentials $q^{a}$. The integrability of pseudopotentials
$q^{a}$ requires that the ideal generated by the form sets $\{ \alpha^{a} \}$
and $\{ \Omega^{a} \}$ is closed, {\it i.\,e.},
\begin{equation}
\label{e2.4}
d\Omega^{a} = \eta^{a}_{b}\wedge\Omega^{b} + f^{a,i}\alpha_{i},
\end{equation}
where $\eta^{a}_{b}$ and $f^{a,i}$ are some 1-forms and 0-forms respectively.
When (\ref{e2.4}) is explicitly written out by using (\ref{e2.2}) and
(\ref{e2.3}), it splits up into a set of partial differential equations:
\begin{equation}
\label{e2.5}
\left.
\begin{array}{l}
\partial_{\phi}F^{a}=0, \,\,\,\,\,\,\,\,\,\,\,\,\,\,
\partial_{\pi_{\phi}}G^{a}=0, \\
F^{b}\partial_{b}G^{a}- G^{b}\partial_{b}F^{a} +
\pi_{\phi} \partial_{\phi}G^{a}
- 2 (e^{\phi} - e^{-2\phi}) \partial_{\pi_{\phi}}F^{a} =0,
\end{array}
\right.
\end{equation}
where the derivative ${\displaystyle \frac{\partial}{\partial q^{a}} }$ is
abbreviated to $\partial_{a}$. Analyzing these equations we find,
\begin{equation}
\label{e2.6}
F^{a}=X^{a}_{0} + X^{a}_{1}\pi_{\phi}, \,\,\,\,\,\,\,\,\,
G^{a}=2Y^{a}_{0}e^{\phi} +2Y^{a}_{1}e^{-2\phi}.
\end{equation}
In the ansatz (\ref{e2.6}) $X^{a}_{i}$ and $Y^{a}_{i}$ $(i=0, \, 1)$ are
assumed to be functions of $q^{a}$ only.

For the convenience of the later discussions we now introduce some vector
fields (Lie derivatives) $X_{i}$ and $Y_{i}$ in $N$-dimensional space of
pseudopotentials ($q$-space),
\begin{equation}
\label{e2.7}
X_{i}=X^{a}_{i}\partial_{a}, \,\,\,\,\,\,\,\,\,\,\,\,\,\,\,\,
Y_{i}=Y^{a}_{i}\partial_{a}.
\end{equation}
It is then a direct consequence of (\ref{e2.5})-(\ref{e2.7}) that,
\begin{equation}
\label{e2.8}
[\,X_{0},\,Y_{0}\,] = X_{1} ,
\,\,\,  [\,X_{0},\,Y_{1}\,] = -X_{1} ,
\,\,\, [\,X_{1},\,Y_{0}\,] = -Y_{0} ,
\,\,\, [\,X_{1},\,Y_{1}\,] = 2Y_{1}.
\end{equation}
Because of the absence of $[\,X_{0},\,X_{1}\,]$ and $[\,Y_{0},\,Y_{1}\,]$,
the set of Lie brackets given by (\ref{e2.8}) does not form a closed
linear algebra. It is obviously impossible that one may close the algebra
by setting the unknown commutators to be linear combinations of the
given generators such that the results are consistent with the Jacobi
identities. This is a big difference between the prolongation structure
of ZMS equation and those of sine-Gordon equation, Ernst equation and
chiral model\cite{Omote,LG}. However, the algebra can be closed by
assigning new generators to the unknown commutators and repeating the
process of working through the Jacobi identities. After a tedious but
straightforward computation, we see that the enlarged algebra becomes
an infinite-dimensional algebra $a^{(2)}_{2}$ (without center), which
coincides with a well-known fact that ZMS model is the Toda field theory
over the twisted Kac-Moody algebra $a^{(2)}_{2}$\cite{Dodd,ZS}.
Let $\{H^{(m)}_{i}, \, E^{(m)}_{\pm \alpha} \,\}$ be the Cartan-Weyl basis
of $a^{(2)}_{2}$, which obey the following commutation relations:
\begin{eqnarray}
\label{e2.9}
&  &[\,H^{(m)}_{i},\,E^{(n)}_{\pm\beta}\,] = \pm \delta_{i1}\delta_{\beta1}
      E^{(m+n)}_{\pm 1} \mp 2\delta_{i1}\delta_{\beta2} E^{(m+n)}_{\pm 2}
      \mp \delta_{i1}\delta_{\beta 3} E^{(m+n)}_{\pm 3}  \nonumber\\
&  &\,\,\,\,\,\,\,\,\,\,\,\,\,\,\,\,\,\,\,\,\,\,\,\,\,\,\,\,\,\,\,\,\,\,
     \,\,\,\,\,\,\,\pm 3\delta_{i2} \delta_{\beta1} E^{(m+n)}_{\mp 3}
     \mp 3 \delta_{i2} \delta_{\beta3} E^{(m+n)}_{\mp 1} \,, \nonumber \\
&  &[\,E^{(m)}_{\pm \alpha},\,E^{(n)}_{\pm \beta}\,] =  \mp \delta_{\alpha 1}
      \delta_{\beta 2} E^{(m+n)}_{\pm 3} \pm \delta_{\alpha 1}
      \delta_{\beta 3} H^{(m+n)}_{2} \, ,  \\
&  &[\,E^{(m)}_{\alpha},\,E^{(n)}_{-\beta}\,] = ( \delta_{\alpha 1}
      \delta_{\beta 1} - \delta_{\alpha 2} \delta_{\beta 2} -
      \delta_{\alpha 3} \delta_{\beta 3} ) H^{(m+n)}_{1} - \delta_{\alpha 3}
      \delta_{\beta 2} E^{(m+n)}_{1} \,  \nonumber \\
&  &\,\,\,\,\,\,\,\,\,\,\,\,\,\,\,\,\,\,\,\,\,\,\,\,\,\,\,\,\,\,\,\,\,\,\,\,
      \,\,\,\,\,- \delta_{\alpha 2} \delta_{\beta 3} E^{(m+n)}_{- 1}
       + 2 \delta_{\alpha 3} \delta_{\beta 1} E^{(m+n)}_{2}
       + 2 \delta_{\alpha 1} \delta_{\beta 3} E^{(m+n)}_{-2} \, ,\nonumber
\end{eqnarray}
where $(m, \, n =0,\,\pm1,\,\pm2,\,\pm3,\, \cdots; \,\,i=1,\,2; \,\,\alpha,
\, \beta =1,\,2,\,3.)$. Then we have the following identifications,
\begin{equation}
\label{e2.10}
X_{0}= E^{(-1)}_{1} + E^{(0)}_{2} ,\,\,\,\, X_{1}= H^{(0)}_{1},\,\,\,\,
Y_{0}= E^{(1)}_{-1},\,\,\,\, Y_{1}= E^{(0)}_{-2} .
\end{equation}

Now we study the linear realizations of the vector fields $X_{i}$ and
$Y_{i}$ in an infinite-dimensional $q$-space which has coordinate
variables $\{q^{(m)}_{j};\, j=1,\,2,\,3;\,m=0,\,\pm1,\,\pm2,\,\pm3,\,
\cdots\}$. Following Omote\cite{Omote}, we introduce some auxiliary vector
fields $\{A^{(m)}_{ij}\}$:
\begin{equation}
\label{e2.11}
A^{(m)}_{ij}=\sum\limits_{n=-\infty}^{+\infty}q^{(m+n)}_{i}\frac{\partial}
{\partial{q^{(n)}_{j}}}.
\end{equation}
They can be shown to satisfy commutator relations
\begin{equation}
\label{e2.12}
[\,A^{(m)}_{ij},\,A^{(n)}_{kl}\,]=\delta_{jk}A^{(m+n)}_{il} -
\delta_{il}A^{(m+n)}_{kj}.
\end{equation}
This fact implies that the set of vector fields $\{A^{(m)}_{ij}\}$ provide an
operator version of graded matrices $\{e^{(m)}_{ij}=e_{ij}\otimes\lambda^{m}
\}$ ($\lambda$ is a gradation parameter). Therefore, the Cartan-Weyl basis of
$a^{(2)}_{2}$ under consideration has the following linear
realization\cite{YCS} in a triplicated infinite-dimensional $q$-space:
$$
\left.\begin{array}{ll}
{\left\{
\begin{array}{l}
H^{(m)}_{1}=A^{(m)}_{11}-A^{(m)}_{33}, \,\, \\
H^{(m)}_{2}=A^{(m)}_{11}-2A^{(m)}_{22} + A^{(m)}_{33}, \,\,\\
E^{(m)}_{2}=A^{(m)}_{31},\,\,\\
E^{(m)}_{-2}=A^{(m)}_{13},\,\,
\end{array}
\right.} &
{\left\{
\begin{array}{l}
E^{(m)}_{1}=A^{(m)}_{12}-A^{(m)}_{23}, \,\,\\
E^{(m)}_{-1}=A^{(m)}_{21}-A^{(m)}_{32},\,\,\\
E^{(m)}_{3}=A^{(m)}_{32}+A^{(m)}_{21}, \,\, \\
E^{(m)}_{-3}=A^{(m)}_{12}+A^{(m)}_{23}.
\end{array}
\right.}
\end{array}
\right.
$$
Explicitly,
\begin{equation}
\label{e2.13}
\left\{
\begin{array}{l}
X_{0}=\sum\limits^{+\infty}_{n=-\infty} \left[ q^{(n-1)}_{1}{\displaystyle
\frac{\partial}{\partial{q^{(n)}_{2}}} } - q^{(n-1)}_{2} {\displaystyle
\frac{\partial}{\partial{q^{(n)}_{3}}} } + q^{(n)}_{3} {\displaystyle
\frac{\partial}{\partial{q^{(n)}_{1}}} } \right] , \\
X_{1}=\sum\limits^{+\infty}_{n=-\infty} \left[ q^{(n)}_{1} {\displaystyle
\frac{\partial}{\partial{q^{(n)}_{1}}} } - q^{(n)}_{3} {\displaystyle
\frac{\partial}{\partial{q^{(n)}_{3}}} } \right] ,\\
Y_{0}=\sum\limits^{+\infty}_{n=-\infty} \left[ q^{(n+1)}_{2} {\displaystyle
\frac{\partial}{\partial{q^{(n)}_{1}}} } - q^{(n+1)}_{3} {\displaystyle
\frac{\partial}{\partial{q^{(n)}_{2}}} } \right] ,  \\
Y_{1}=\sum\limits^{+\infty}_{n=-\infty} \left[ q^{(n)}_{1} {\displaystyle
\frac{\partial}{\partial{q^{(n)}_{3}}} } \right] ,
\end{array}
\right.
\end{equation}
$i.\,e.$, the components of $F^{a}$ and $G^{a}$ are assigned to the following
linear representations in a triplicated infinite-dimensional $q$-space :
\begin{equation}
\label{e2.14}
\left. \begin{array}{lll}
F^{(n)}_{1}=q^{(n)}_{3} + q^{(n)}_{1} \pi_{\phi} ,\,\,\,\,
&F^{(n)}_{2}=q^{(n-1)}_{1} ,\,\,\,\,
&F^{(n)}_{3}=-q^{(n-1)}_{2} - q^{(n)}_{3} \pi_{\phi} ,\\
G^{(n)}_{1}=2q^{(n+1)}_{2} e^{\phi}  ,\,\,\,\,
&G^{(n)}_{2}=-2q^{(n+1)}_{3} e^{\phi} ,\,\,\,\,
&G^{(n)}_{3}=2q^{(n)}_{1} e^{-2\phi}.
\end{array}
\right.
\end{equation}

Relying on the 1-form (\ref{e2.3}), we see that the pseudopotentials
introduced in (\ref{e2.14}) satisfy equations
\begin{eqnarray}
\label{e2.15}
&  &\partial_{-}{\left [
\begin{array}{l}
q^{(n)}_{1} \\ q^{(n)}_{2} \\ q^{(n)}_{3}
\end{array}
\right ] }
={\left [
\begin{array}{lll}
\partial_{-}\phi \,\, &\,\,0 \,\, &\,\,\,\,1 \,\,\\
\,\,0 \,\, &\,\,0 \,\, &\,\,\,\,0 \,\,\\
\,\,0 \,\, &-1 \,\, &-\partial_{-}\phi \,\,
\end{array}
\right ] }
{\left [
\begin{array}{l}
q^{(n)}_{1} \\ q^{(n-1)}_{2} \\ q^{(n)}_{3}
\end{array}
\right ] }
+{\left [
\begin{array}{lll}
0 \,\, &0 \,\, &0 \,\, \\
1 \,\, &0 \,\, &0 \,\, \\
0 \,\, &0 \,\, &0 \,\,
\end{array}
\right ] }
{\left [
\begin{array}{l}
q^{(n-1)}_{1} \\ q^{(n)}_{2} \\ q^{(n)}_{3}
\end{array}
\right]}, \nonumber \\
&  &\partial_{+}{\left [
\begin{array}{l}
q^{(n)}_{1} \\ q^{(n)}_{2} \\ q^{(n)}_{3}
\end{array}
\right ] }
={\left [
\begin{array}{lll}
\,\,\,\,0 \,\, &2e^{\phi} \,\, &\,\,\,\,0 \,\,\\
\,\,\,\,0 \,\, &\,\,\,\,0 \,\, &-2e^{\phi} \,\,\\
2e^{-2\phi} \,\, &\,\,\,\,0 \,\, &\,\,\,\,0 \,\,
\end{array}
\right ] }
{\left [
\begin{array}{l}
q^{(n)}_{1} \\ q^{(n+1)}_{2} \\ q^{(n+1)}_{3}
\end{array}
\right ] },
\end{eqnarray}
on the solution surface $(\phi(x^{+},\,x^{-}),\,\pi_{\phi}(x^{+},\,x^{-}),\,
x^{+},\,x^{-})$ of ZMS Eq.(\ref{e2.1}). Let us define a parameter-dependent
potential $\Psi(\lambda)$ by
\begin{equation}
\label{e2.16}
\Psi(\lambda) \equiv {\left [
\begin{array}{l}
\psi_{1}(\lambda) \\ \psi_{2}(\lambda) \\ \psi_{3}(\lambda)
\end{array}
\right ] } = \sum\limits^{+\infty}_{n=-\infty} \lambda^{n} \,
{\left [
\begin{array}{l}
q^{(n)}_{1} \\ q^{(n)}_{2} \\ q^{(n)}_{3}
\end{array}
\right ] }.
\end{equation}
Then we get from (\ref{e2.15}) the partial differential equations for
$\Psi(\lambda)$: \,$\partial_{\pm}\Psi = A_{\pm}\Psi$, where
\begin{equation}
\label{e2.17}
\left\{ \begin{array}{l}
A_{+} = {\displaystyle \frac{2}{\lambda} } e^{\phi} E_{1}
+ 2 e^{- 2\phi} E_{2} , \\
A_{-} = \partial_{-}\phi H_{1} + \lambda E_{- 1} + E_{- 2} .
\end{array}
\right.
\end{equation}
Such $A_{\pm}$ do just constitute a Lax pair representation of ZMS
Eq.(\ref{e2.1}), which gives Eq.(\ref{e2.1}) as the zero-curvature
equation $[\partial_{+} - A_{+}, \, \partial_{-} - A_{-}] = 0$, the
consistency condition of the equations for $\Psi$. In (\ref{e2.17}), $H_{1},
\, E_{\pm 1}$ and $E_{\pm 2}$ are some $3 \times 3$ matrices defined as
$H_{1}= e_{11}-e_{33}, \, E_{1}=e_{12}-e_{23}, \, E_{2}=e_{31}, \,
E_{-1}=e_{21}-e_{32}$ and $E_{-2}=e_{13}$ respectively. These matrices are
among the independent generators of $SL(3,{\cal R})$ group\cite{YCS}.

Another aspect of the prolongation structure is the nonlinear realizations
of the vector fields $X_{i}$ and $Y_{i}$\, $(i=0,\,1)$ in a
finite-dimensional $q$-space. In Refs.\cite{Omote,LG}, the authors
cited some instances in illustration of the fact that the linear realizations
of the vector fields are relevant to Lax representation while the nonlinear
realizations of these fields associate themselves with the B$\ddot{a}$cklund
transformation of the considered nonlinear differential equation. We will
show that this conclusion is also true for ZMS Eq.(\ref{e2.1}).

It is worthwhile to indicate that there does not exist an unpenetrable
barrier between the linear realizations and the nonlinear realizations
of the vector fields. In fact, there is a standard method to
get the nonlinear realizations of the vector fields from their linear
realizations, in which the Lax pair plays the crucial role\cite{Kaup}.
Let us apply the method to ZMS model. Defining $q_{1}=\psi_{1}/\psi_{2},\,
q_{2}=-\psi_{3}/\psi_{2}$ as new pseudopotentials, we see from Lax pair
(\ref{e2.17}) and auxiliary linear equations that these
pseudopotentials\cite{Nucci}
satisfy a set of Riccati-type equations:
\begin{equation}
\label{e2.18}
\left. \begin{array}{ll}
\partial_{+}q_{1}={\displaystyle \frac{2}{\lambda} } (1 - q_{1} q_{2})
e^{\phi} , \,\,
&\partial_{+}q_{2}= - 2 q_{1} e^{-2\phi}  -
{\displaystyle \frac{2}{\lambda} } q^{2}_{2} e^{\phi} , \\
\partial_{-}q_{1}= - (q_{2} + \lambda q^{2}_{1}) + q_{1} \pi_{\phi}, \,\,
&\partial_{-}q_{2}=\lambda (1 - q_{1} q_{2} ) - q_{2} \pi_{\phi} ,
\end{array}
\right.
\end{equation}
where $\pi_{\phi}=\partial_{-}\phi$. On the other hand, it follows from
(\ref{e2.3}) and (\ref{e2.6}) that
\begin{equation}
\label{e2.19}
\left. \begin{array}{ll}
\partial_{+}q_{1}=2Y^{1}_{0} e^{\phi} + 2Y^{1}_{1} e^{-2\phi}, \,\,
&\partial_{+}q_{2}= 2Y^{2}_{0} e^{\phi} + 2Y^{2}_{1} e^{-2\phi}, \\
\partial_{-}q_{1}= X^{1}_{0} + X^{1}_{1}\pi_{\phi} , \,\,
&\partial_{-}q_{2}= X^{2}_{0} + X^{2}_{1}\pi_{\phi} ,
\end{array}
\right.
\end{equation}
in two-dimensional $q$-space. Therefore, the vector fields acquire the
following nonlinear realizations:
\begin{equation}
\label{e2.20}
\left\{
 \begin{array}{l}
X_{0}= - (q_{2} + \lambda q^{2}_{1} )\partial_{1}
+ \lambda (1 - q_{1} q_{2} )\partial_{2} , \\
X_{1}= q_{1} \partial_{1} - q_{2} \partial_{2} , \\
Y_{0}= {\displaystyle \frac{1}{\lambda} } (1 - q_{1} q_{2} )\partial_{1} -
{\displaystyle \frac{1}{\lambda} } q^{2}_{2} \partial_{2} , \\
Y_{1}= - q_{1} \partial_{2} .
\end{array}
\right.
\end{equation}
Simultaneously, the prolongated algebra $a^{(2)}_{2}$ is nonlinearly
realized as,
\begin{equation}
\label{e2.21}
\left.\begin{array}{ll}
{\left\{
\begin{array}{l}
H^{(m)}_{1}=\lambda^{-m} ( q_{1} \partial_{1} - q_{2} \partial_{2}) ,\,\,\\
H^{(m)}_{2}=3\lambda^{-m} ( q_{1} \partial_{1} + q_{2} \partial_{2}) ,\,\,\\
E^{(m)}_{2}= - \lambda^{-m} q_{2} \partial_{1} , \,\,\\
E^{(m)}_{-2}= - \lambda^{-m} q_{1} \partial_{2} , \,\,
\end{array}
\right.}
& {\left\{
\begin{array}{l}
E^{(m)}_{1}=-\lambda^{-m} [ q^{2}_{1} \partial_{1} -(1 - q_{1} q_{2} )
\partial_{2} ], \,\,\\
E^{(m)}_{-1}= \lambda^{-m} [ (1 - q_{1} q_{2} ) \partial_{1}
- q^{2}_{2} \partial_{2}] , \,\, \\
E^{(m)}_{3}=\lambda^{-m} [ (1 + q_{1} q_{2} ) \partial_{1}
+ q^{2}_{2} \partial_{2}] , \,\, \\
E^{(m)}_{-3}=-\lambda^{-m} [ q^{2}_{1} \partial_{1} + (1 + q_{1} q_{2} )
\partial_{2} ].
\end{array}
\right.}
\end{array}
\right.
\end{equation}
In (\ref{e2.20}) and (\ref{e2.21}), \,$\lambda$ is an arbitrary spectral
parameter and $m$ takes integer value. Notably, the vector fields
(\ref{e2.20}) are not among a finite-dimensional subalgebra of $a^{(2)}_{2}$
unless $\lambda$ equals to {\it one}. This is another important difference in
the prolongation structure of ZMS equation from those of sine-Gordon
equation, Ernst equation and chiral model\cite{Omote}.

\section{The B$\ddot{a}$cklund transformation of ZMS equation}

The aim of this section is to search the auto-B$\ddot{a}$cklund
transformation of ZMS Eq.(\ref{e2.1}) on the basis of the prolongation
structure. We assume that the new ZMS field variables $\widetilde{\phi}$
and $\widetilde{\pi_{\phi}}$ are functions of the old $\phi$ ,\,
$\pi_{\phi}$ and the pseudopotentials $q^{a}$. The new forms $\alpha_{i}\,
(i=1,\,2)$ which are got from the old ones by replacing $\phi$ and
$\pi_{\phi}$ with $\widetilde{\phi}(\phi,\,\pi_{\phi},\,q^{a})$ and
$\widetilde{\pi_{\phi}}(\phi,\,\pi_{\phi},\,q^{a})$ should vanish modulo
the old $\alpha_{i}$ and the prolongation 1-forms $\Omega^{a}$; $i.\,e.$,
there should exist some 0-forms $g_{ij}$ and 1-forms $\nu_{i}^{a}$ such
that
\begin{equation}
\label{e3.1}
\widetilde{\alpha}_{i}(\widetilde{\phi},\,\widetilde{\pi_{\phi}}) = g_{ij}
\alpha_{j}(\phi,\,\pi_{\phi}) + \nu_{i}^{a} \wedge \Omega^{a}.
\end{equation}
This is the condition for the existence of B$\ddot{a}$cklund transformations.
In terms of (\ref{e2.2}) and (\ref{e2.3}), the condition can be recast as
\begin{equation}
\label{e3.2}
\partial_{\pi_{\phi}} \widetilde{\phi} = 0, \,\,\,\,\,\,
\partial_{\phi} \widetilde{\pi_{\phi}} = 0,\,\,\,\,\,\,
\widetilde{\pi_{\phi}} = \pi_{\phi} ( \partial_{\phi} + X_{1} )
\widetilde{\phi} + X_{0} \widetilde{\phi},
\end{equation}
and
\begin{equation}
\label{e3.3}
e^{\widetilde{\phi}} - e^{-2 \widetilde{\phi}} =
( e^{\phi} - e^{-2 \phi} ) \partial_{\pi_{\phi}} \widetilde{\pi_{\phi}} +
( e^{\phi} Y_{0} + e^{-2\phi} Y_{1} ) \widetilde{\pi_{\phi}} .
\end{equation}
The special expressions of Eqs.(\ref{e3.2}) lead to the following ansatz
solutions for the new ZMS field variables $\widetilde{\phi}$ and
$\widetilde{\pi_{\phi}}$:
\begin{equation}
\label{e3.4}
\left\{ \begin{array}{l}
\widetilde{\phi} = c \phi + f(q^{a}), \\
\widetilde{\pi_{\phi}} = \pi_{\phi} [c + X_{1}f(q^{a})] + X_{0}f(q^{a}) ,
\end{array}
\right.
\end{equation}
where $c$ is an outstanding constant. Substituting (\ref{e3.4}) into
(\ref{e3.3}) we get,
\begin{equation}
\label{e3.5}
X_{1}f(q^{a}) = c_{1},
\end{equation}
and,
\begin{equation}
\label{e3.6}
e^{ c \phi} e^{f(q^{a})} - e^{-2 c \phi } e^{-2 f(q^{a})} =  e^{\phi} [\,c
+ c_{1} + Y_{0}X_{0} f(q^{a}) \,] - e^{-2 \phi} [\, c + c_{1}
- Y_{1}X_{0} f(q^{a}) \,] .
\end{equation}
where $c_{1}$ is another constant.

We now apply the nonlinear expressions (\ref{e2.20}) of the vector fields
$X_{i}$ and $Y_{i}$\, $(i=0,\,1)$ in the two-dimensional $q$-space to
Eqs.(\ref{e3.4})-(\ref{e3.6}). In this case, (\ref{e3.5}) becomes a
first-order quasi-linear differential equation whose general solution reads:
\begin{equation}
\label{e3.7}
f(q_{1},\,q_{2}) = c_{1} \ln q_{1} + \omega(q_{1} q_{2}) ,
\end{equation}
where $\omega(q_{1} q_{2})$ is an arbitrary differentiable function of its
variable $q_{1} q_{2}$. Generally speaking, Eq.(\ref{e3.5}) would have
another particular solution beyond (\ref{e3.7}). But we quit finding such
a particular solution here. One of the reason is that there is no systematic
method for searching it. What is more, even if we happened to find out a
particular solution for Eq.(\ref{e3.5}), it would be excessive to expect
this solution satisfying Eq.(\ref{e3.6}) further. If the B$\ddot{a}$cklund
transformation of ZMS equation exists, it is bound to connect with the
general solution (\ref{e3.7}).

To determine the function $\omega(q_{1} q_{2})$ in (\ref{e3.7}) and then
resolve completely the B$\ddot{a}$cklund transformation for ZMS
Eq.(\ref{e2.1}), Eq.(\ref{e3.6}) must be taken into account. After a simple
and straightforward calculation we find,
\begin{equation}
\label{e3.8}
c=1, \,\,\,\,c_{1}=0,\,\,\,\,\omega(q_{1} q_{2}) = \ln ( 2 q_{1} q_{2} - 1 ).
\end{equation}
Namely, the auto-B$\ddot{a}$cklund transformation of ZMS Eq.(\ref{e2.1})
is as follows:
\begin{equation}
\label{e3.9}
\widetilde{\phi} =  \phi + \ln ( 2 q_{1} q_{2} - 1 ),
\end{equation}
where the auxiliary pseudopotentials are determined by Riccati equations
\begin{equation}
\label{e3.10}
\left. \begin{array}{ll}
\partial_{+}q_{1}={\displaystyle \frac{2}{\lambda} } (1 - q_{1} q_{2})
e^{\phi} , \,\,
&\partial_{+}q_{2}= - 2 q_{1} e^{-2\phi}  -
{\displaystyle \frac{2}{\lambda} } q^{2}_{2} e^{\phi} , \\
\partial_{-}q_{1}= - (q_{2} + \lambda q^{2}_{1})
+ q_{1} \partial_{-}\phi, \,\,
&\partial_{-}q_{2}=\lambda (1 - q_{1} q_{2} ) - q_{2}\partial_{-}\phi ,
\end{array}
\right.
\end{equation}
for a given $\phi$.
One can easily justify that $\widetilde{\phi}$ is a solution
of ZMS Eq.(\ref{e2.1}) once $\phi$ fulfills Eq.(\ref{e2.1}) and vice versa.
The problem of B\"{a}cklund transformation of ZMS equation was
ever discussed in
\cite{Sharipov,Safin}. The B\"{a}cklund transformation obtained
by Sharipov and Yamilov \cite{Sharipov} is a set of second order
differential equations, which is very cumbersome for solving
single soliton solutions\cite{Safin} and is difficult to turn
to our B\"{a}cklund transformation (\ref{e3.9}-\ref{e3.10}).

\section{Discussions}

In the previous sections we have studied the prolongation structure of
Zhiber-Mikhailov-Shabat equation. The prolongation structure yields an
incomplete set of commutators of vector fields in the pseudopotential
space. Following Omote\cite{Omote} and Ablowitz $et \,al$\cite{Kaup}, we
have found out the linear and nonlinear differential realizations of the
vector fields respectively. It is shown that the linear realizations of
the vector fields give the linear auxiliary equations (Lax pair) of ZMS
equation, while their nonlinear realizations are connected with the
B$\ddot{a}$cklund transformation of the equation. Nevertheless, the
application of this B$\ddot{a}$cklund transformation to constructing new
analytical solutions of ZMS equation from some old ones, {\it e.g.},
constructing the single-soliton solution from vacuum solution, remains
an open problem. It is easy to see that ZMS Eq.(\ref{e2.1}) has an analytical
solution governed by the first-order equations $\partial_{\pm}\phi =
\mu^{\pm} \sqrt{2(2e^{\phi} + e^{-2\phi} - 3)}$ ($\mu$ is an arbitrary
constant). But we have not driven out these equations from the
B$\ddot{a}$cklund transformation laws yet. Another even more interesting
problem is perhaps to study the dressing group symmetry in ZMS model by
virtue of prolongation structure. By some naive perception we conjecture
that the finite dressing transformations of ZMS equation may be exposed
through a slightly different nonlinear realization of the prolongation
structure. The dressing procedure is a useful method for solving
single soliton solution.
The detail for such problems are now in preparation.

\acknowledgments

This work is partially supported by NSF of Zhejiang province. H.X. Yang
acknowledge the support of the Science Foundation for Postdoctoral
Research in China.


\vskip 3cm
\begin{thebibliography}{99}
\bibitem{Dodd}
R.K.Dodd, R.K.Bullough, Proc. Roy. Aoc. Lond. {\bf A352}, 481(1977).
\bibitem{ZS}
A.V.Zhiber and A.B. Shabat, Dokl. Akad. Nauk USSR {\bf 247}, No.5(1979).
\bibitem{Lamb}
G.L.Jr Lamb, Rev. Mod. Phys. {\bf 43}, 99(1971).
\bibitem{Mikhailov}
A.V.Mikhailov, Pisma v ZhETF {\bf 30}, 443(1979).
\bibitem{Korepin}
A.G.Izergin and V.E.Korepin, Commun. Math. Phys. {\bf 79}, 303(1981).
\bibitem{Smirnov}
F.A.Smirnov, Int. J. Mod. Phys. {\bf A6}, 1407(1991).
\bibitem{Efthimiou}
C.J.Efthimiou, Nucl. Phys. {\bf B398}, 697(1993).
\bibitem{YCS}
Y.X.Chen, H.X.Yang and Z.M.Sheng, Phys. Lett. {\bf B345}, 149(1995).
\bibitem{WE}
D.H.Wahlquist and F.B.Estabrook, J. Math. Phys. {\bf 16}, 1(1975).
\bibitem{Omote}
M.Omote, J. Math. Phys. {\bf 27}, 2853(1986).
\bibitem{LG}
Y.Q.Lee and M.L.Ge, Commun. Theor. Phys. {\bf 10}, 79(1988).
\bibitem{Kaup}
M.J.Ablowitz, D.J.Kaup, A.C.Newell and H.Segur, Stud. Appl. Math.
{\bf 53}, 249(1974).
\bibitem{Nucci}
M.C.Nucci, Riccati-type pseudopotentials and their applications in
"Nonlinear Equations in the Applied Sciences", ed W.F.Ames and
Colin Rogers, Academic Press (1992).
\bibitem{Sharipov}
R.A.Sharipov and R.I.Yamilov, "Backlund transformation and the
construction of an integrable boundary-value problem for the equation
$u_{xt} = e^u - e^{-2u}$" in: Problems of Mathematical Physics
and Asymptotic Behavior of Their Solitons [in Russian]
(Ufa 1991).
\bibitem{Safin}
S.S.Safin and R.A.Sharipov,
Theoretical Mathematical Physics, {\bf 94},462(1993).
\end{thebibliography}
\end{document}